\documentclass[%
 reprint,
 amsmath,amssymb,
 aps,
 pre,
]{revtex4-2}

\usepackage[colorlinks=true,allcolors=blue]{hyperref}
\usepackage{graphicx} 
\usepackage{bm}
\usepackage{amsmath, amssymb}
\usepackage{braket}
\usepackage{xcolor}

\newcommand{\Figref}[1]{Fig.~\ref{#1}}
\newcommand{\Figsref}[1]{Figs.~\ref{#1}}
\newcommand{\Eqref}[1]{~(\ref{#1})}

\newcommand{\tk}[1]{{\color{black}{#1}}}
\usepackage{ulem}
\begin{document}


\title{
Propulsion of a chiral swimmer in viscoelastic fluids
}


\author{Takuya Kobayashi}
\email{kobayashi@cheme.kyoto-u.ac.jp}
\author{John J. Molina}
\author{Ryoichi Yamamoto}
\email{ryoichi@cheme.kyoto-u.ac.jp}
\affiliation{Department of Chemical Engineering, Kyoto University, Kyoto 615-8510, Japan}


\date{\today}

\begin{abstract}
Microswimmers often use chirality to generate translational movement from rotation motion, exhibiting distinct behaviors in complex fluids compared to simple Newtonian fluids. However, the underlying mechanism remains incompletely understood. In this study, we elucidate the precise mechanisms underlying the distinct behaviors of microswimmers in Newtonian and non-Newtonian fluids. We show that the enhanced speed of chiral swimmers is attributed to the Weissenberg effect induced by normal stress differences resulting from chiral flows. Additionally, we identify swimmer-specific normal stress differences in a viscoelastic fluid and demonstrate that swimming speed varies depending on whether the swimmer acts as a pusher or a puller. \tk{Moreover, we investigate the hydrodynamic interactions between a pair of chiral squirmers. When the squirmers are aligned parallel (perpendicular) to their swimming axis, they tend to separate (approach).}
These findings deepen our comprehension of the rheological properties of viscoelastic fluids containing microswimmers, promising advancements in various applications.
\end{abstract}


\maketitle



%


\section{Introduction}
The rheological characteristics of complex fluids often deviate from well-explored Newtonian fluids. A notable example of this deviation is the Weissenberg effect, where viscoelastic fluids climb up a rotating rod due to normal stress differences~\cite{Weissenberg1947-lx}.
In various biological scenarios, self-propelled cells often interact with complex fluids, exhibiting non-Newtonian behavior. Living biological microswimmers in complex fluids include bacteria forming viscoelastic biofilms~\cite{Donlan2002-pg, Sillankorva2012-xh} and sperm cells navigating viscoelastic mucus toward the ovum for fertilization~\cite{Katz1978-nj, fauci2006biofluidmechanics, Suarez2006-yg}.
Similarly, artificial chiral microswimmers have been engineered to navigate solely in complex fluids whereas they fail to propel themselves in Newtonian fluids~\cite{Pak2012-rk, Puente-Velazquez2019-fs, Rogowski2021-no, Binagia2021-fi, Kroo2022-qm}. 
It is well-established that microswimmers behave differently in complex fluids compared to Newtonian fluids~\cite{Patteson2015-pa, Patteson2016-mf, Tung2017-po, Li2021-fv}.
For instance, {\it E. coli} cells swim faster and tumble less often in both polymer solutions~\cite{Patteson2015-pa} and colloidal suspensions~\cite{Kamdar2022-id}, compared to Newtonian fluids; sperm cells swim collectively in polymer solutions, but not in Newtonian fluids~\cite{Tung2017-po}. Previous experimental, theoretical and simulation studies have shown that fluid elasticity can affect the swimming speeds of microswimmers in complex viscoelastic fluids. This is observed for a wide variety of swimmers, including filaments~\cite{Fu2007-ks, Fu2009-jy}, sheets~\cite{Lauga2007-pi, Teran2010-fj, Riley2014-cl}, undulatory swimmers~\cite{Shen2011-cm, Binagia2019-bv}, helical flagella~\cite{Liu2011-rw, Spagnolie2013-tj} and squirmers~\cite{Zhu2011-eo, Zhu2012-wp, De_Corato2015-nq, Datt2019-ym, Binagia2020-dn, Housiadas2021-qe, Kobayashi2023-bs}.
Understanding microswimming in complex fluids holds relevance for biological and medical sciences, as well as various health applications, such as biofilm control, fertilization, and drug delivery systems~\cite{Patra2013-lm, Gao2014-xt}.

{\it E. coli} cells exhibit counterclockwise flagellar rotation and clockwise body rotation during swimming, resulting in chiral flow~\cite{Lauga2016-zq}, and they swim faster in polymer solutions compared to Newtonian fluids~\cite{Patteson2015-pa}. Theoretical and simulation works have indicated that the swimming speed of squirmers is enhanced due to the coupling between chirality and viscoelasticity~\cite{Binagia2020-dn, Housiadas2021-qe}. Experimental, theoretical, and simulation studies have recently shown that micro-robots can swim in viscoelastic fluids by applying this same principle~\cite{Binagia2021-fi, Kroo2022-qm}. A chiral micro-robot, consisting of the counter-rotating head and tail spheres, could swim in viscoelastic fluids but not in Newtonian fluids~\cite{Binagia2021-fi, Kroo2022-qm}. Previous studies highlight the critical role of the first normal stress difference in boosting swimming speeds~\cite{Pak2012-rk, Puente-Velazquez2019-fs, Rogowski2021-no, Binagia2020-dn, Binagia2021-fi, Housiadas2021-qe, Kroo2022-qm}. However, the precise mechanisms underlying the generation of normal stress differences (NSD) by the swimmers' chirality and their impact on swimming speed remain unclear. 
In this study, we expand on previous works~\cite{Binagia2020-dn, Housiadas2021-qe, Kobayashi2023-bs}, investigating the generation of the first NSD by the swimmer's chirality, resulting in the Weissenberg effect, which is responsible for the swimming speed enhancement.

Furthermore, previous theoretical and simulation studies~\cite{Zhu2012-wp, De_Corato2015-nq, Datt2019-ym, Kobayashi2023-bs}, using the squirmer model, have shown that the swimming speed varies with the swimmer type (pusher, puller or neutral) in viscoelastic fluids. In contrast, in Newtonian fluids at low Reynolds numbers, the swimming speed is relatively insensitive to the swimmer's type~\cite{Wang2012-sl}. The mechanisms responsible for the swimmer-type dependence of the swimming speed in various viscoelastic fluids remain to be determined. In this study, we elucidate that the first NSD is also responsible for the swimmer-type dependence of the swimming speed. 

To understand the influence of the fluid elasticity on the swimming speed, we conducted direct numerical simulations of a squirmer in an Oldroyd-B fluid, using the Smoothed Profile (SP) method~\cite{Nakayama2005-aa, Yamamoto2021-oe} to fully resolve the hydrodynamic coupling between the fluid flow and the squirmer particle.
We have clarified the precise mechanism behind (i) the acceleration in swimming speed due to the swimmer's chirality and (ii) variations in swimming speed based on swimmer type in viscoelastic fluids.
To understand the variation in the swimming speed between Newtonian and viscoelastic fluids, we calculated the force on a fixed squirmer and analyzed the normal stress differences around the particle.
\tk{Additionally, we investigated how squirmers hydrodynamically interact with each other in an Oldroyd-B fluid and measured the interaction forces between chiral squirmers. We found that the squirmers' chirality influences their hydrodynamic interactions. Specifically, when squirmers are aligned parallel or perpendicular to their swimming axes, they tend to move away from or towards each other, respectively.}
Our findings will not only advance the applications of drug delivery systems for artificial micro-robots but also provide valuable insights into natural microswimming in complex fluid environments, such as fertilization of the chiral corkscrew-shaped movement of frog sperm cells~\cite{Muto2009-zj}.

\section{Simulation Methods}
\subsection{Squirmer model}
\begin{figure}
    \centering
    \includegraphics[width=0.9\linewidth]{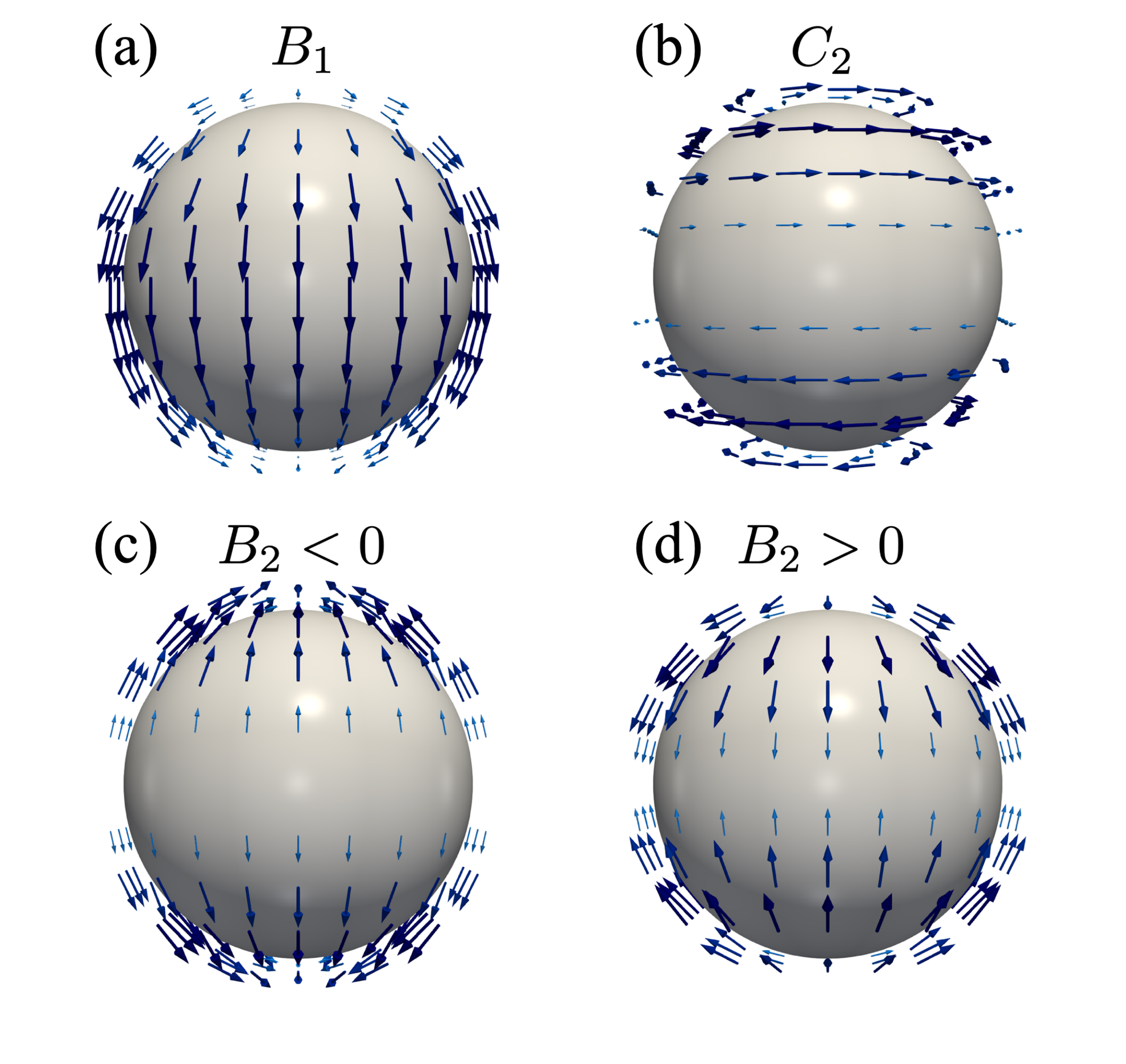}
    \caption{Schematic representation of the surface slip velocity of each mode, (a)$B_1$, (c, d)$B_2$, and (b)$C_2$.}
    \label{fig:squirmer}
\end{figure}
We represent microswimmers using the rigid spherical squirmer model developed by Lighthill~\cite{Lighthill1952-ui} and Blake~\cite{Blake1971-ws} to represent ciliary propulsion.
In the squirmer model, the propulsion is generated by imposing a modified slip velocity at the particle's surface, which takes the general form~\cite{Pak2014-uf}:
\begin{align}
    \bm{u}^{\rm sq}(\theta, \varphi) &= \sum_{n = 1}^\infty \frac{2}{n(n + 1)}B_nP_n'(\cos\theta)\sin\theta\hat{\bm{\theta}}\notag\\ &\ + \sum_{n = 1}^\infty C_nP_n'(\cos\theta)\sin\theta\hat{\bm{\varphi}}
\end{align}
where $P'_n$ is the derivative of the $n$-th order Legendre polynomial, $B_n$ and $C_n$ are the coefficients for the $n$-th polar and azimuthal squirming modes, respectively, with $\hat{\bm{\theta}}$ and $\hat{\bm{\varphi}}$ the corresponding unit tangent vectors.
Specifically, we consider the first two polar modes $B_1$ and $B_2$, along with the second azimuthal mode $C_2$, the so-called rotlet dipole. This consideration leads to the following expression for the slip velocity (\Figref{fig:squirmer}):
\begin{align}
    \bm{u}^{\rm sq}(\theta, \varphi) &= \left(B_1\sin\theta + \frac{B_2}{2}\sin2\theta\right)\hat{\bm{\theta}} + \frac{3}{2}C_2\sin2\theta\hat{\bm{\varphi}}\label{eq:slip vel}\\
    &= B_1\left[\left(\sin\theta + \frac{\alpha}{2}\sin2\theta\right)\hat{\bm{\theta}} + \frac{3}{2}\zeta\sin2\theta\hat{\bm{\varphi}}\right]
\end{align}
The first polar mode $B_1$ determines the steady swimming speed in a Newtonian fluid, expressed as $U_{\rm N} = 2/3B_1$. Furthermore, the ratios $\alpha = B_2 / B_1$ and $\zeta = C_2 / B_1$ represent the swimmer type and the strength of the flow chirality, respectively. Swimmers with $\alpha <0 $ are called pushers (e.g., {\it E.coli.}), which generate an extensile flow field in the swimming direction. On the other hand, $ \alpha >0 $ denotes pullers (e.g., {\it Chlamydomonas}), which generate a contractile flow field. Squirmers with $\alpha = 0$ are termed neutral swimmers, and swim with a potential flow field. The chiral parameter $\zeta$ generates the flow pattern of a microswimmer with rotating flagella and a counter-rotating body~\cite{Fadda2020-ve,Binagia2020-dn, Housiadas2021-qe}.

\subsection{Smoothed Profile Method}
We employed the Smoothed Profile (SP) method \cite{Nakayama2005-aa, Yamamoto2021-oe} to fully resolve the fluid-particle hydrodynamic coupling. This method has been employed to investigate the dynamics of colloids and microswimmers in viscoelastic fluids~\cite{matsuoka_2020, matsuoka_2021, Kobayashi2023-bs}. In this method, the sharp boundary between the particle and fluid is replaced with a continuous diffuse interface of thickness $\xi$, by introducing a smoothed profile function $\phi\in[0, \ 1]$, which is 1 inside the particle and 0 outside. The mathematical definition of $\phi$ can be found in Ref.~\cite{Nakayama2005-aa}. Introducing this $\phi$ field allows us to define field variables for all particle quantities. The fluid-particle coupling is then accounted for by enforcing the momentum conservation between fluid and particle domains. In particular, the total velocity $\bm{u} = \bm{u}_{\rm f} + \bm{u}_{\rm p}$, including the host fluid $\bm{u}_{\rm f}$ and particle contributions $\bm{u}_{\rm p}$, is governed by a modified Navier-Stokes equation
\begin{align}
    \rho\left(\frac{\partial}{\partial t}+\bm{u}\cdot\bm{\nabla}\right)\bm{u} = -\bm{\nabla}p + \bm{\nabla}\cdot\bm{\sigma} + \rho(\phi\bm{f}_{\rm p} + \phi\bm{f}_{\rm sq})
\end{align}
with the incompressible condition $\bm{\nabla}\cdot\bm{u} = 0$, where $\rho$ is the fluid mass density, $p$ is the pressure, $\bm{\sigma}$ is the stress tensor, $\phi\bm{f}_{\rm p}$ is a body force applied to enforce the particle's rigidity, and $\phi\bm{f}_{\rm sq}$ is the force due to the squirming motion. For the viscoelastic fluid, we adopt the Oldroyd-B model ($\bm{\sigma} = \bm{\sigma}_{\rm s} + \bm{\sigma}_{\rm p}$), which is a generalization of the Newtonian fluid, $\bm{\sigma}_{\rm s} = \eta_{\rm s}\left[\bm{\nabla}\bm{u} + \left(\bm{\nabla}\bm{u}\right)^{\rm T}\right]$, and the Upper Convected Maxwell (UCM) fluid, $\bm{\sigma}_{\rm p} + \lambda \stackrel{\triangledown}{\bm{\sigma}}_{\rm p}= \eta_{\rm p}\left[\bm{\nabla}\bm{u} + \left(\bm{\nabla}\bm{u}\right)^{\rm T}\right]$,
where $\eta_{\rm s}$ and $\eta_{\rm p}$ are the viscosities of the Newtonian solvent and the viscoelastic solute, respectively, $\lambda$ is the polymer relaxation time, and $\stackrel{\triangledown}{\bm{\sigma}}$ is the upper-convected derivative, given by
$
    \stackrel{\triangledown}{\bm{\sigma}} = \partial_t \bm{\sigma} + \bm{u}\cdot\bm{\nabla}\bm{\sigma} - (\bm{\nabla}\bm{u})^{\rm T} \cdot\bm{\sigma} - \bm{\sigma}\cdot(\bm{\nabla}\bm{u})
$.

The (spherical) particles follow the Newton-Euler equations of motion, given by
\begin{align}
    \dot{\bm{R}}_i &= \bm{V}_i,\qquad
    \quad\dot{\bm{Q}}_i={\rm skew}(\bm{\Omega}_i)\cdot\bm{Q}_i\\
    M_i\dot{\bm{V}}_i &= \bm{F}_i^{\rm H} + \bm{F}_i^{\rm sq},\qquad
    \bm{I}_{\rm p}\cdot\dot{\bm{\Omega}}_i = \bm{N}_i^{\rm H} + \bm{N}_i^{\rm sq}
\end{align}
 where $\bm{R}_i$ and $\bm{V}_i$ are the position and velocity of particle $i$, respectively, and $\bm{Q}_i$ and $\bm{\Omega}_i$ the orientation matrix and angular velocity. The hydrodynamic forces and torques are denoted as $\bm{F}_i^{\rm H}$ and $\bm{N}_i^{\rm H}$, respectively, while $\bm{F}_i^{\rm sq}$ and $\bm{N}_i^{\rm sq}$ represent the force and torque due to the squirming motion. The function ${\rm skew}(\bm{\Omega}_i)$ is used to create the skew-symmetric matrix of the angular velocity $\bm{\Omega}_i$, which is defined as:
\begin{align}
    {\rm skew}(\bm{\Omega}) = 
     \left(\begin{array}{ccc}
         0&-\Omega_z&\Omega_y  \\
         \Omega_z&0&-\Omega_x\\
         -\Omega_y & \Omega_x & 0
    \end{array}
    \right)
\end{align}
Further details can be found in Ref.~\cite{Kobayashi2023-bs}.

\section{Swimming speed of a single squirmer}
\subsection{Chirality dependence}\label{chiral_speed}
\begin{figure}
    \centering
    \includegraphics[width=0.9\linewidth]{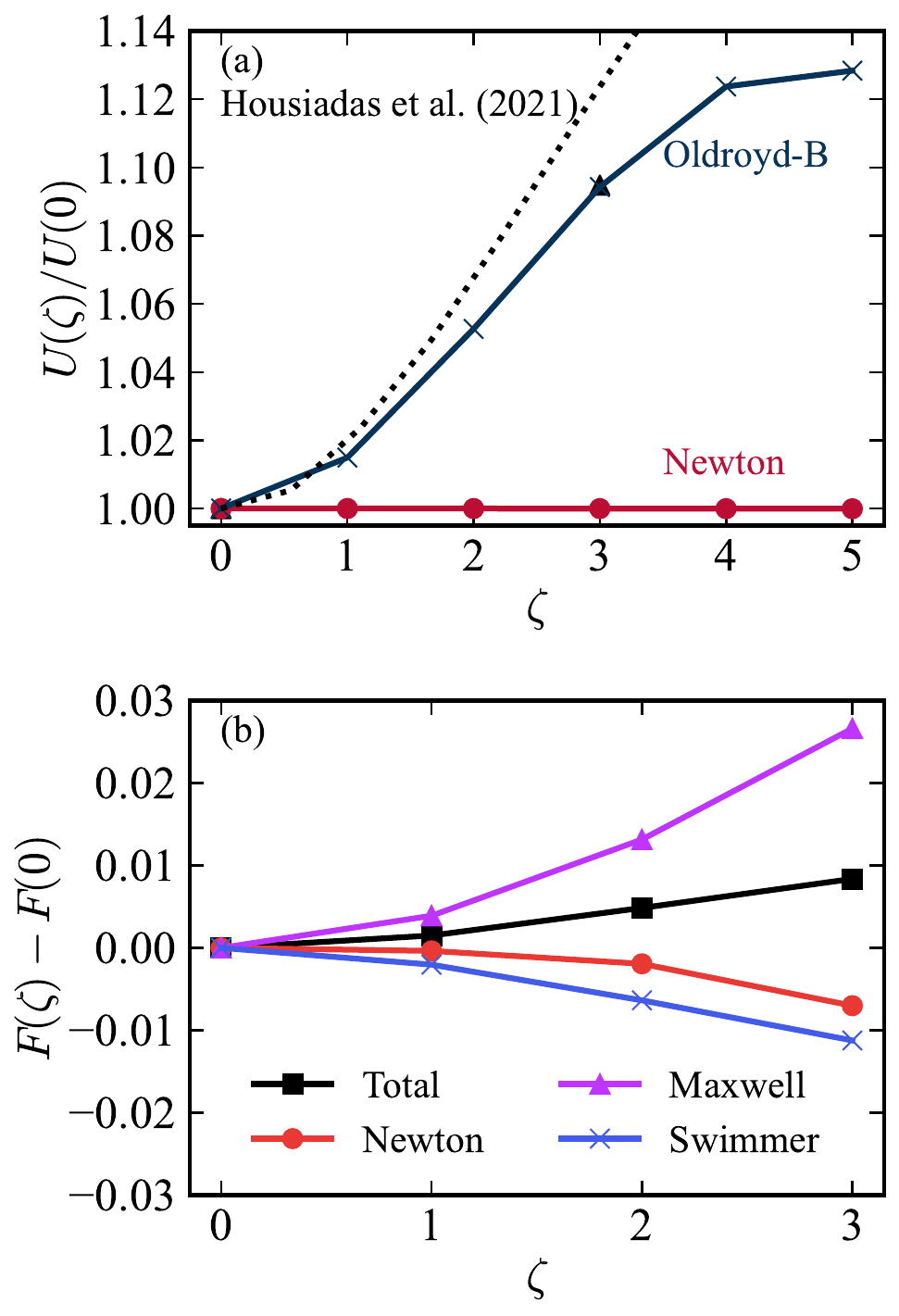}
    \caption{
    (a) Nondimensional swimming speed of {a single neutral squirmer under a periodic boundary condition} as a function of the chiral parameter $\zeta$ in a Newtonian fluid (red) and an Oldroyd-B fluid (blue). The symbols represent simulation results. The dotted line and triangle symbols are the analytical solution and numerical results obtained by Housiadas {\it et al.}~\cite{Housiadas2021-qe} in an Oldroyd-B fluid, respectively.
    (b) Force difference from the force at $\zeta = 0$ on a fixed squirmer in an Oldroyd-B fluid, as a function of the chiral parameter $\zeta$ with the force being parallel to the swimming axis.
    }
    \label{fig:rotlet:speed_force}
\end{figure}

We performed simulations for a single neutral squirmer, radius $a = 5\Delta$ and interface width $\xi = 2\Delta$, inside a cubic simulation box of length $L = 128\Delta$ ($\Delta$ represents the grid spacing), with periodic boundary conditions in all directions. The host fluids have a mass density and zero-shear viscosity of $\rho=\eta_0 = 1$. The particle Reynolds number, Weissenberg number and the viscosity ratio are ${\rm Re} = \rho U_{\rm N} a /\eta_0 = 0.005$, ${\rm Wi} = \lambda B_1 / a = 0.2$ and $\beta = \eta_{\rm s} / \eta_0 = \eta_{\rm s} / (\eta_{\rm s} + \eta_{\rm p}) = 0.5$, respectively. 
It is observed that the swimming velocity in viscoelastic fluids increases with the chiral parameter $\zeta$ (see \Figref{fig:rotlet:speed_force}(a)). 
This speed-up aligns with reports for Gisekus fluid~\cite{Binagia2020-dn} and Oldroyd-B fluids~\cite{Housiadas2021-qe}, demonstrating good agreement with the numerical results in Oldroyd-B fluids~\cite{Housiadas2021-qe}. {This previous study has also reported an analytical solution for the swimming speed in Oldroyd-B fluids, with a quartic increase in the degree of chirality ($U(\zeta) / U(0) \sim \zeta^4$)~\cite{Housiadas2021-qe}}, which is in quantitative agreement with simulation results, at least for small $\zeta$.
In contrast, the velocity remains constant in Newtonian fluids, regardless of $\zeta$, since the $C_2$ term is decoupled from the translational motion for a spherical particle. 
Previously, we found that, due to the squirmer's chirality, the enhancement occurs for all swimmer types (pusher, puller or neutral) in Oldroyd-B fluids~\cite{Kobayashi2023-bs}.

In order to unveil the mechanism behind this swimming speed enhancement, we shift our focus to the force contributions parallel to the swimming axis on a fixed squirmer. As shown in \Figref{fig:rotlet:speed_force}(b), the total force increases with the chiral parameter $\zeta$, similar to the swimming speed. 
Furthermore, the force contributions arising from the Newtonian stress $\bm{\sigma}_{\rm s}$ and squirming motion both decrease with the chiral parameter $\zeta$. In contrast, the contribution due to the Maxwell stress $\bm{\sigma}_{\rm p}$ increases with the chiral parameter $\zeta$. Therefore, we can conclude that the Maxwell contribution is the primary mechanism responsible for the swimming speed enhancement.

\begin{figure}[tb!]
    \centering
    \includegraphics[width=\linewidth]{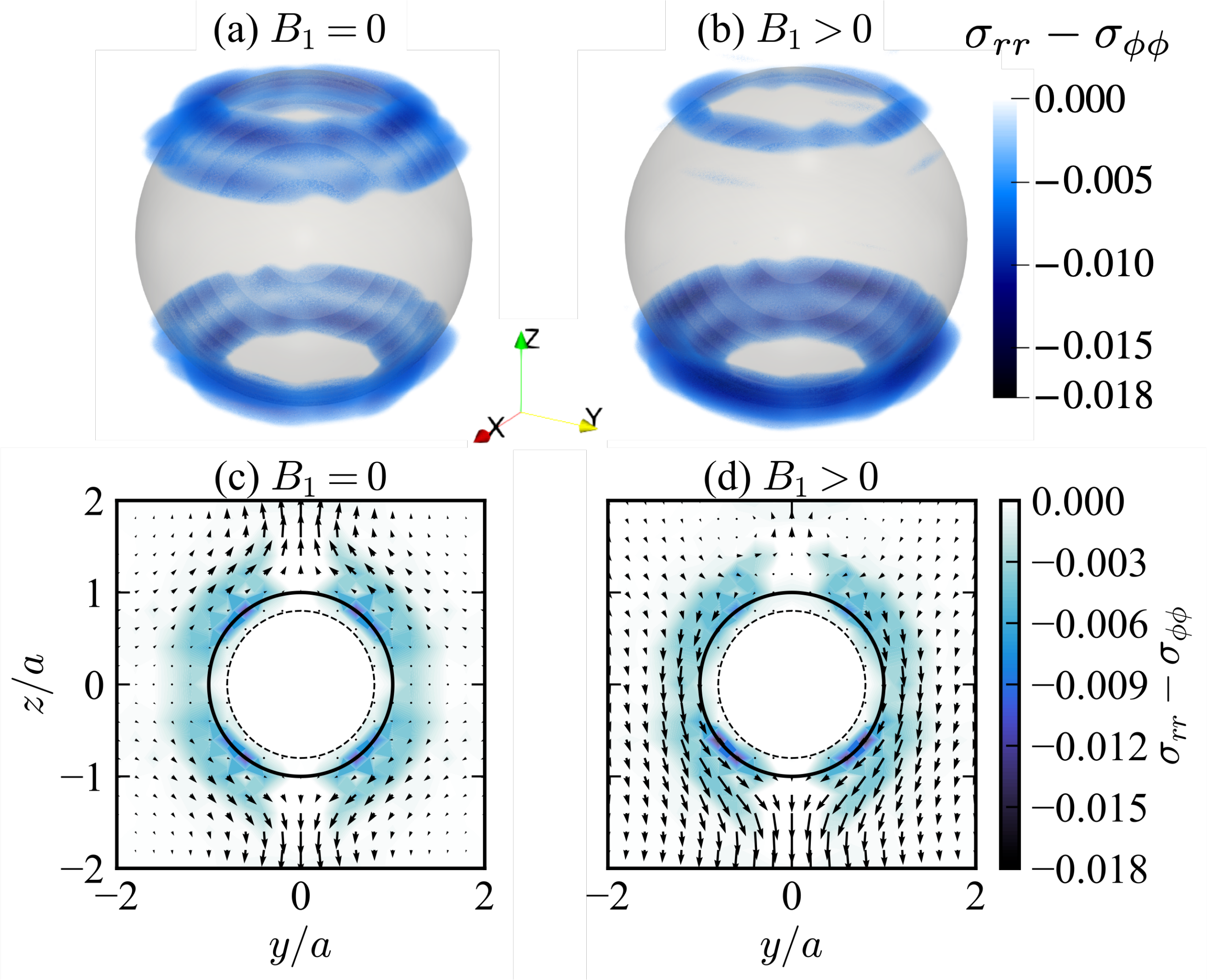}
    \caption{Velocity and first normal stress difference (NSD) $\sigma_{rr} - \sigma_{\phi\phi}$ fields around a squirmer particle. (a) and (c) correspond to a squirmer with $B_1 = 0$, while (b) and (d) correspond to $B_1 > 0$. In (c) and (d), the quiver plot shows the velocity field, and the density map represents the NSD field. The dashed lines represent the position of $r = a - \xi / 2$.
    }
    \label{fig:rotlet:Weissenberg}
\end{figure}

{To expand upon the previous works~\cite{Binagia2020-dn, Housiadas2021-qe, Kobayashi2023-bs}, and clarify how the Maxwell stress contributes to the total force, we considered the role of the normal stress differences (NSD), which have been previously reported to enhance the swimming speed~\cite{Binagia2020-dn, Housiadas2021-qe, Kroo2022-qm}.} It is worth noting that the main distinction between Newtonian and Oldroyd-B fluids lies in the presence of the first NSD.
The force per unit area on a squirmer, in spherical coordinates, is given by the first NSD
\begin{align}
    \left(\bm{\sigma}_{\rm f}\cdot \bm{n}\right)_r = \sigma_{rr} - p= \sigma_{rr} - \sigma_{\phi\phi} + {\rm const.}
\end{align}
By considering only the rotlet dipole ($C_2$ term) in Eq.~\Eqref{eq:slip vel} and taking into account rotational symmetry in the $\phi$ direction, $\partial_\phi (-p + \sigma_{\phi\phi})$ = 0. 

As shown in \Figref{fig:rotlet:Weissenberg}, this first NSD $\sigma_{rr} - \sigma_{\phi\phi}$, generated by the squirmer's chirality, adopts a ring-shaped distribution near the north/south poles. In turn, this results in a net force directed towards the center of the squirmer.
When $B_1 = 0$, the NSD field is symmetric between the upper and lower hemispheres. Thus, the NSD works towards reducing/enhancing the swimming speed in the upper/lower hemispheres, however, given the symmetry, these two contributions balance each other and result in a net zero total force.
However, when $B_1 \ne 0$, the asymmetry of the velocity field between the upper and lower hemispheres, generated by the asymmetry of the $B_1$ term (as described in Eq.~\Eqref{eq:slip vel}), leads to a top-bottom asymmetry in the normal stress differences. In particular, the absolute value of the NSD on the lower hemisphere is greater than that on the upper hemisphere. The NSD thus works towards reducing/enhancing the swimming speed in the upper/lower hemispheres, with the force on the lower hemisphere being greater, which leads to the swimming speed enhancement in viscoelastic fluids.

{
In previous work~\cite{Housiadas2021-qe}, the forces acting on a freely swimming squirmer were calculated in force- and torque-free conditions, but the detailed mechanism behind the swimming speed enhancement of a chiral squirmer in Oldroyd-B fluids was not elucidated. Building on this, we elucidate the detailed mechanism behind the swimming speed enhancement of a chiral squirmer in Oldroyd-B fluids. Specifically, we compute the force acting on a fixed squirmer, for which the force and torque are balanced by the external force required to fix it, and examine the normal stress differences around it. \Figref{fig:rotlet:Weissenberg} illustrates the Weissenberg effect resulting from the counter-rotating flows in the upper and lower hemispheres. When $B_1 > 0$, the top-bottom asymmetric Weissenberg effect generates a greater force on the lower hemisphere compared to the upper hemisphere, thereby accelerating the squirmer in Oldroyd-B fluids.}

\subsection{Swimmer type dependence}
\begin{figure}
    \centering
    \includegraphics[width=0.9\linewidth]{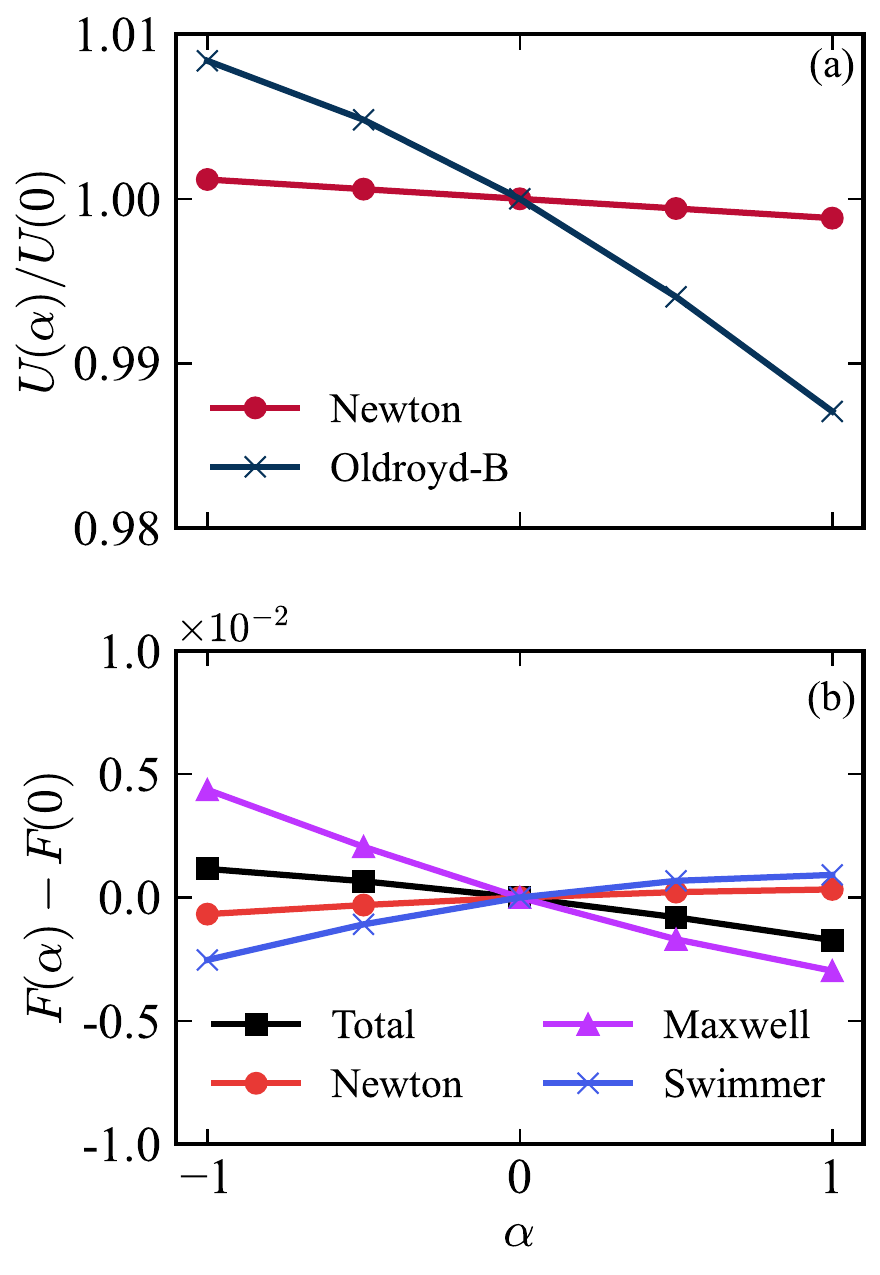}
    \caption{
    (a) Nondimensional swimming speed of {a single squirmer under a periodic boundary condition} as a function of the swimming type $\alpha$ in both a Newtonian fluid (red) and an Oldroyd-B fluid (blue).
    (b) Force difference from the force at $\alpha = 0$ on a fixed squirmer in an Oldroyd-B fluid as a function of the swimmer type $\alpha$, with the force being parallel to the swimming axis.
    }
    \label{fig:st:speed_force}
\end{figure}

\begin{figure}[htb!]
    \centering
    \includegraphics[width=\linewidth]{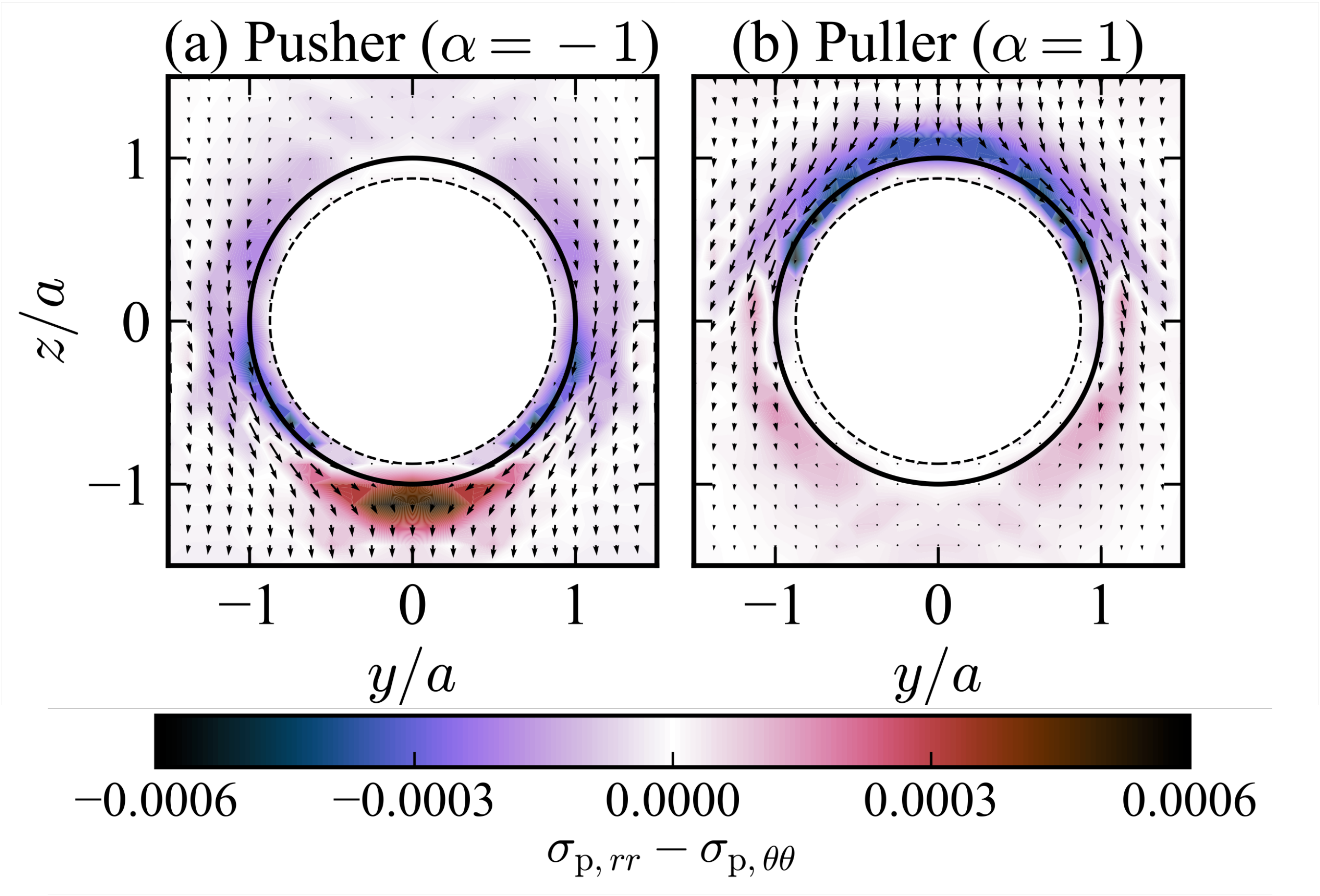}
    \caption{Velocity field and first normal stress difference $\sigma_{rr} - \sigma_{\theta\theta}$ for (a) a pusher with $\alpha = -1$ and (b) a puller with $\alpha=1$. The quiver plot shows the velocity field and the density map represents the NSD field. The dashed lines represent the position of $r = a - \xi / 2$.}
    \label{fig:st:NSD}
\end{figure}
Next, we conducted simulations for a single non-chiral squirmer ($\zeta = 0$), with $a = 8\Delta$ and $\xi = 2\Delta$ within a fully periodic cubic simulation box of length $L = 128\Delta$.
The particle Reynolds number, Weissenberg number, and viscosity ratio are ${\rm Re} = 0.005$, ${\rm Wi} =  0.1$, and $\beta = 0.5$, respectively. 
The swimming speeds for a squirmer as a function of the swimmer type $\alpha$ are shown in \Figref{fig:st:speed_force}(a). In a Newtonian fluid, the swimming speed is independent of the swimming type, with a week Re dependence, which can be approximated by $U / U_{\rm N} \simeq 1 - 0.15\alpha{\rm Re}$~\cite{Wang2012-sl}. This is consistent with our observed results. 
In contrast, in an Oldroyd-B fluid, the swimming speed of a squirmer decreases with $\alpha$, with pushers ($\alpha < 0$) swimming faster than pullers ($\alpha > 0$). 
This observation closely resembles the perturbation expansion in terms of $\rm Wi$ for the UCM fluid (an Oldroyd-B fluid with $\eta_{\rm s} = 0$), which is expressed as $U / U_{\rm N} = 1 - 0.2\alpha{\rm Wi}$~\cite{Datt2019-ym}. Remarkably, this trend holds despite the difference in the fluids (i.e., constitutive relations).

We again analyzed the force contributions parallel to the swimming axis on a fixed squirmer. As shown in \Figref{fig:st:speed_force}(b), the force decreases with $\alpha$, as does the swimming speed. We find that
pushers swim faster than pullers due to the decrease in the Maxwell contribution with the $\alpha$ parameter. To clarify this decrease in the Maxwell contribution to the force with $\alpha$, we investigated the NSD around a pusher with $\alpha = -1$ and a puller with $\alpha = 1$.
As illustrated in \Figref{fig:st:NSD}, 
for a pusher, the NSD due to uniaxial elongation at $\theta = \pi$ works towards reducing the swimming speed, while the NSD due to the shear flow at $\theta \simeq 3\pi / 4$ can enhance the swimming speed. On the other hand, for a puller, both the NSD due to biaxial elongation at $\theta = 0$ and the NSD due to the shear flow at $\theta \simeq \pi/4$ work towards reducing the swimming speed. 
Therefore, the swimming speed dependence on the swimming type $\alpha$ is associated with the direction of the force generated by the shear flow in an Oldroyd-B fluid, while it is related to the force generated by the elongational flow in a Giesekus fluid~\cite{Zhu2012-wp}.

\section{Hydrodynamic interactions}
\begin{figure*}[tb]
    \centering
    \includegraphics[width=0.95\linewidth]{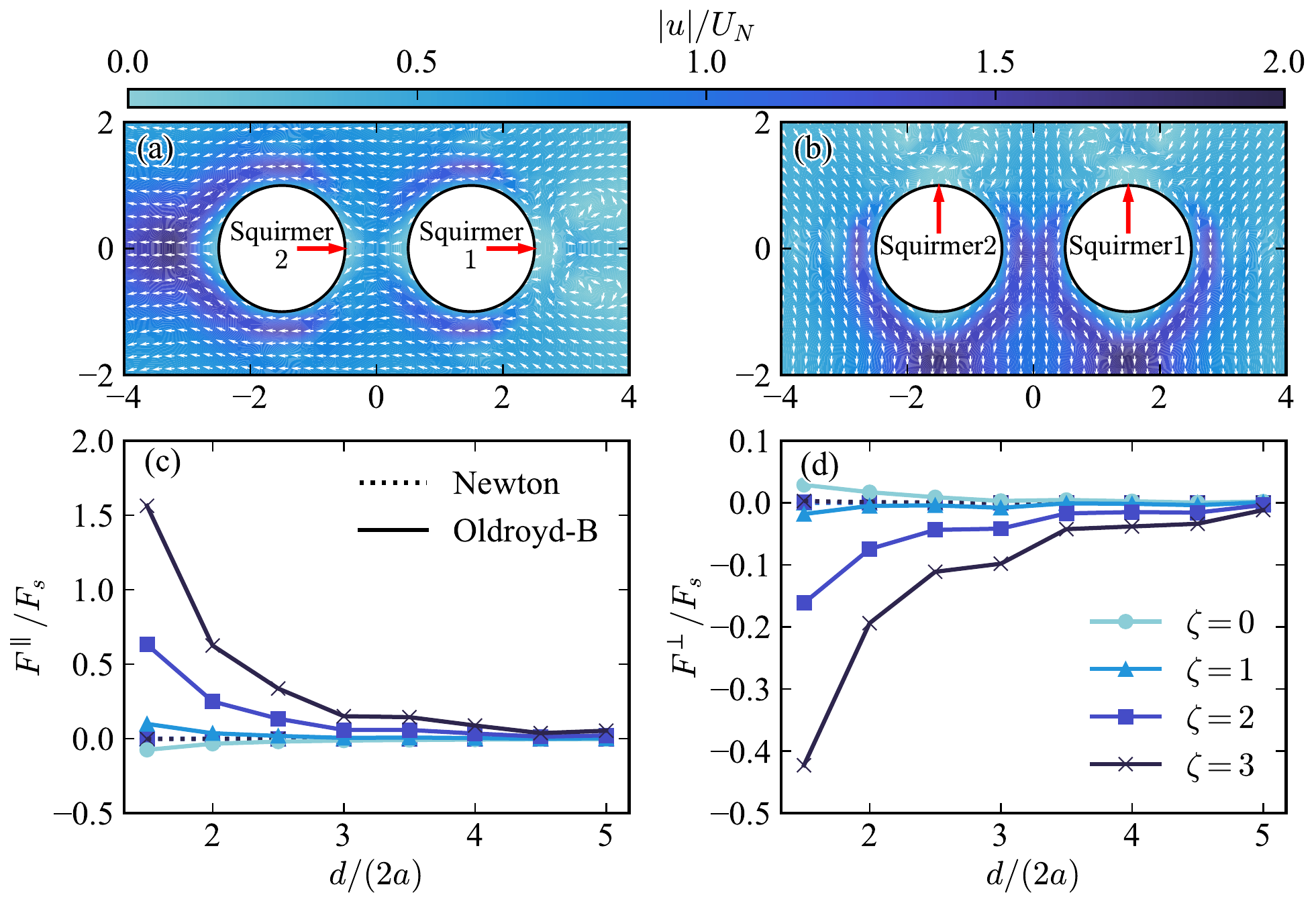}
    \caption{
    (a), (b) Velocity fields around two fixed neutral squirmers with $\zeta = 3$ in the steady state for a Weissenberg number $\rm Wi = 0.2$ and a viscosity ratio $\beta = 0.5$. The distance between the two squirmers is $d / (2a) = 1.5$ in the directions (a) parallel and (b) perpendicular to the swimming axis. The colormaps depict the absolute values of the fluid velocity, normalized by the Newtonian swimming speed $U_N$.
    (c), (d) Hydrodynamic interaction forces between two squirmers fixed at a distance $r$ in the directions (c) parallel and (d) perpendicular to the swimming axis in the steady state. The forces are scaled by the Stokes force at the Newtonian swimming speed $F_s = 6\pi\eta a U_N$. The dotted and solid lines represent the force in Newtonian and Oldroyd-B fluids, respectively.
    }
    \label{fig:HI}
\end{figure*}

In addition to examining hydrodynamic squirmer interactions in Newtonian fluids~\cite{Ishikawa2006-hw, Gotze2010-by}, we investigate the hydrodynamic interactions between two fixed squirmers with parallel orientations in Oldroyd-B fluids. The separation distance between the two squirmers is $d$, both parallel and perpendicular to the swimming axis. Our simulations involved two neutral squirmers with a radius $a = 5\Delta$ and an interface width $\xi = 2\Delta$. These were placed inside a cubic simulation box with a length $L = 128\Delta$, using periodic boundary conditions in all three directions. The host fluids have a mass density and zero-shear viscosity of $\rho = \eta_0 = 1$. $B_1$ was chosen to satisfy the condition that ${\rm Re} = \rho U_N a/ \eta_0 = 0.005$. The Weissenberg number and the viscosity ratio were set to ${\rm Wi} = 0.2$ and $\beta = 0.5$, respectively. 

The resulting velocity fields are illustrated in \Figsref{fig:HI}(a) and (b). The distances between the squirmers are $d/ (2a) = 1.5$ in both the parallel (\Figref{fig:HI}(a)) and perpendicular (\Figref{fig:HI}(b)) directions relative to their swimming axes. In \Figref{fig:HI}(a), squirmer 2 reduces the flow velocity behind squirmer 1. In \Figref{fig:HI}(b), a divergent flow is observed between the squirmers. This flow characteristic is similar to the behavior seen between pushers in Newtonian fluids~\cite{Gotze2010-by} because the chiral squirmer produces pusher-like flow patterns in Oldroyd-B fluids~\cite{Kobayashi2023-bs}, due to the Weissenberg effect induced by squirmer's chirality, as discussed in Sec.~\ref{chiral_speed}.

Furthermore, we calculate the interaction force of fixed parallel-oriented squirmers for various distances $d$ parallel (\Figref{fig:HI}(c)) and perpendicular (\Figref{fig:HI}(d)) to their swimming axes. To understand their tendency to either separate or approach, we calculate the hydrodynamic interaction force $\bm{F} = (\bm{F}_1 - \bm{F}_2)/2$. The force is normalized by the Stokes drag force acting on a sphere moving with the Newtonian swimming velocity $U_{\rm N}$, i.e., $F_s = 6\pi\eta a U_{\rm N}$.
As shown in \Figref{fig:HI}(c), the interaction force between chiral squirmers is positive, meaning they experience a repulsive force. The Weissenberg effect, generated by the chirality of squirmer 1, pushes against squirmer 2, leading to this repulsive force between the squirmers. As a result, squirmer 1 swims faster than squirmer 2 along their swimming axes, causing their separation in Oldroyd-B fluids. 
The force increases with $\zeta$ in Oldroyd-B fluids, while it remains constant, regardless of $\zeta$, in Newtonian fluids.
As illustrated in \Figref{fig:HI}(d), the interaction force between chiral squirmers is negative, indicating an attractive force. Due to the Weissenberg effect caused by the chirality of both squirmers, fluid escapes easily from the gap between them, creating a divergent flow. This results in an attractive force that leads to their mutual approach.
The force decreases with $\zeta$ in Oldroyd-B fluids, while it remains constant, regardless of $\zeta$, in Newtonian fluids.
Depending on whether squirmers are fixed parallel or perpendicular to the swimming axis, the interaction forces typically lead to either separation or approach, respectively, driven by the Weissenberg effect resulting from the squirmer's chirality.

\section{Conclusion}
In conclusion, we investigated the dependence of the swimming speed of a rigid spherical squirmer in Oldroyd-B fluids, on the chiral parameter $\zeta$ and the swimming strength $\alpha$. We used the Smoothed Profile (SP) method to fully resolve the hydrodynamic coupling between the swimmer and the surrounding fluid. To elucidate the mechanism behind the swimming speed, we conducted a detailed analysis of the forces on a fixed squirmer.
We have investigated the precise mechanism behind (i) speed-up due to the swimmer's chirality and (ii) variations in swimming speed based on swimmer type in viscoelastic fluids. 
The first primary mechanism responsible for the speed-up due to the squirmer's chirality is the first NSD, leading to the Weissenberg effect. When $B_1 > 0$, the top-bottom asymmetric Weissenberg effect generates a larger force on the lower hemisphere than on the upper hemisphere, thus accelerating the squirmer in Oldroyd-B fluids.
Furthermore, when considering variations in swimming speed based on swimmer type, the first NSD generates forces that accelerate pushers and decelerate pullers. This explains why pushers swim faster than pullers in Oldroyd-B fluids.
\tk{
Additionally, we investigate the hydrodynamic interactions of chiral squirmers in Oldroyd-B fluids. The interaction forces between a pair of squirmers aligned in a parallel or perpendicular direction to their swimming axes, tend to make them separate or approach each other, respectively, owing to the Weissenberg effect caused by the squirmers' chirality. We anticipate that these distinct hydrodynamic interactions in Oldroyd-B fluids, influenced by the Weissenberg effect, significantly impact the collective behavior of squirmer suspensions in viscoelastic fluids.
}
Our work facilitates the understanding of swimming and flow phenomena in viscoelastic fluids, such as bacteria in biofilms and sperm cells in mucus. Additionally, it should contribute to the design of micro-machines for drug delivery systems.

\begin{acknowledgments}
The authors would like to thank Prof. Takashi Taniguchi for fruitful discussions. This work was supported by the Grants-in-Aid for Scientific Research (JSPS KAKENHI) under Grant Number 20H05619, by the JSPS Core-to-Core Program ``Advanced core-to-core network for the physics of self-organizing active matter (JPJSCCA20230002)'', and by JST SPRING, Grant No. JPMJSP2110.
\end{acknowledgments}


%




\bibliography{manuscript.bbl}


\end{document}